\documentclass[newstyle,twocolumn,proceedings]{rmaa}
\renewcommand{\P}[1]{%
\ifnum#1=1\hbox{OW~168--326E}\fi
\ifnum#1=2\hbox{OW~167--317}\fi
\ifnum#1=3\hbox{OW~163--317}\fi
\ifnum#1=5\hbox{OW~158--323}\fi
\ifnum#1=0\hbox{OW~171--334}\fi}

\title{The Effects of Magnetic Fields on Radiative Cooling Jets}
\author{E.M. de Gouveia Dal Pino\altaffilmark{1} and A. H. 
Cerqueira\altaffilmark{1} 
  \affil{Instituto  Astron\^omico e Geof\'\i sico, IAG-USP, S\~ao Paulo} }

\fulladdresses{
\item E. M. de Gouveia Dal Pino and A. H. Cerqueira: Instituto de Astron\^omico 
e 
Geof\'\i sico,
  Universidade de S\~ao Paulo, A.v Miguel St\'efano 4200, 04301-904 S\~ao Paulo, 
SP, Brazil, (dalpino,adriano@iagusp.usp.br).}

\shortauthor{de Gouveia Dal Pino \& Cerqueira}
\shorttitle{MHD Effects on Cooling Jets}


\abstract{
We here review the results of recent studies of  magnetic field
effects on the structure and evolution of overdense, radiative cooling,
steady and pulsed jets which were carried out with the help of fully
three-dimensional MHD simulations.  Three initial magnetic field
topologies are considered: (i) a helical and (ii) a longitudinal field,
both of which permeate the jet and the ambient medium, and (iii) a
pure toroidal field in the jet. Using  a set of parameters which is
particularly suitable for protostellar jets, we examine the emission
structure of the internal knots and the leading working surface and
compare with a nonmagnetic baseline and also with 2-D MHD calculations.
 }

\listofauthors{E. M. de Gouveia Dal Pino \& A. H. Cerqueira}
\indexauthor{de Gouveia Dal Pino, E.~M.}
\indexauthor{Cerqueira, A.~H.}

\begin{document}

\maketitle

\section{Introduction}
\label{sec:intro}

Low-mass young stellar objects  produce collimated optical outflows
(the  {\it Herbig-Haro}, hereafter HH, jets) that may extend from a few
1000 AU to very large parsec-length scales, and have embedded  in them
bright emission-line knots which are radiating shock fronts propagating
with v $\sim$ few 100 km/s into the ambient medium (e.g., Bally \&
Reipurth 2001, these proceedings).

It has been known for sometime that the dynamics of these jets
is very much affected by radiative cooling due to the recombination
of the shock-excited gas, since the typical cooling times behind the
shocks, which do not exceed a few hundred years, are much smaller
than the dynamical time scales of these jets, $t_{dyn}= R_j/v_j \simeq
10^{4} $ to $10^{5}$ yr (where $R_j$ is the jet radius, and $v_j$ its
velocity). The importance of the cooling on the jet dynamics has been
confirmed by extensive hydrodynamical numerical work (e.g., Blondin et
al. 1990, de Gouveia Dal Pino \& Benz 1993, Stone \& Norman 1993; see also
Reipurth \& Raga 1999, and Cabrit et al. 1997  for reviews).

Another aspect that seem to play a major role both  in the production and
collimation of  the HH jets is the presence of magnetic fields. The most
promising mechanism for  their launching involves magneto-centrifugal
forces associated either with the accretion disk that surrounds the star
(e.g. K\"onigl \& Pudritz 2000), or with the disk-star boundary (in the
X-winds; e.g. Shu et al. 1994).  The details of these launching models
are discussed elsewhere in these proceedings (see, e.g.,  Shang \& Shu
2001, these proceedings). In either case, the poloidal field lines that
are launched from the star-disk system are  expected to wind up like a
spring, thus developing a toroidal component that is able to collimate the
gas outflow towards the rotation axis.  Direct observations of magnetic
fields are difficult, and only recently Ray et al. (1997) have obtained
the first direct evidence, based on polarization measurements, of $B
\sim$ 1 G in the outflow of T Tau S at a distance of few tens of AU from
the source.  Using magnetic flux conservation and typical jet parameters,
this latter result would imply a plasma $\beta \, = \, p_{j,gas}/(B^2/8
\pi) \, \simeq \, 10^{-3}$ for a toroidal field configuration, and $
\sim 10^{3}$, for a longitudinal field, at distances $\sim 0.1$ pc.
Since ambipolar diffusion does not seem to be able to significantly
dissipate them (e.g., Frank et al.  1999), the  figures above suggest
that magnetic fields may also play a relevant role on the outer scales
of the flow.

These peaces of evidence have lately motivated several MHD investigations
of overdense \footnote{Observations suggest that HH jets are denser
than the surrounding ambient medium, with number density ratios $1 <
\eta \gtrsim 10$.}, radiative cooling jets,  in a search for possible
signatures of magnetic fields on the large scales of the HH outflows.
These studies have been carried out with the help of multidimensional
numerical simulations both in two-dimensions (2-D) (Frank et al. 1998,
1999, 2000; Lerry \&  Frank 2000; Gardiner et al.  2000; Stone \&
Hardee 2000; Gardiner \& Frank 2000; O'Sullivan \& Ray 2000), and in
three-dimensions (3-D) (de Gouveia Dal Pino \& Cerqueira 1996; Cerqueira,
de Gouveia Dal Pino \& Herant 1997; Cerqueira \& de Gouveia Dal Pino
1999, 2001a, b). In this lecture, we  try to review and summarize the
main results of these studies.

\begin{figure}
  \includegraphics[width=\columnwidth]{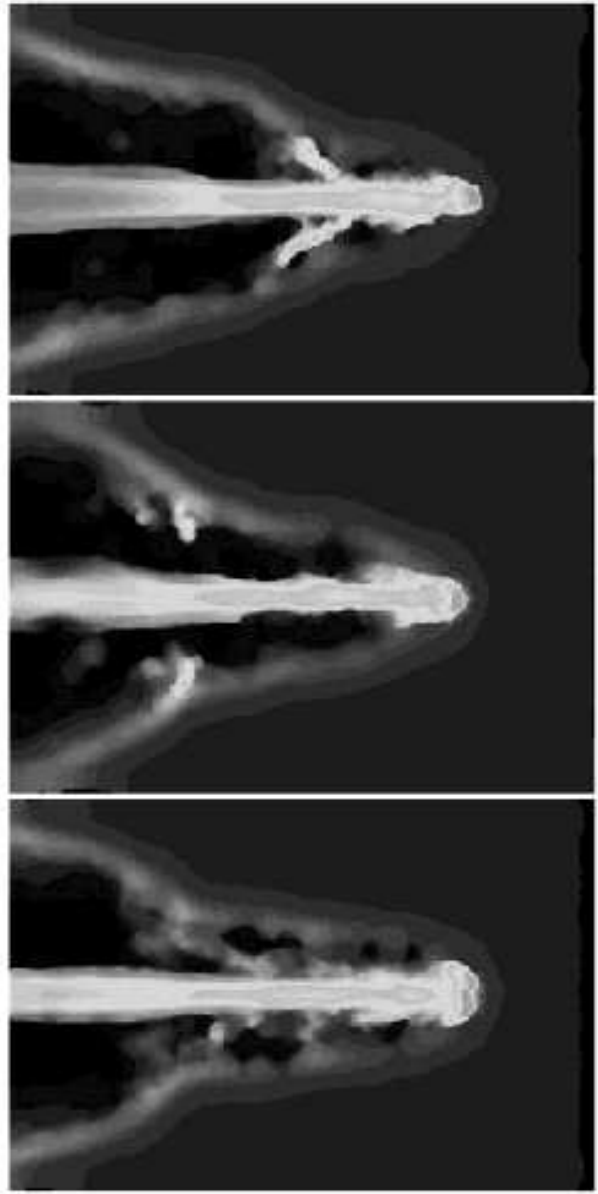}
\caption{Gray-scale representation of the midplane density of the head
of a hydrodynamical jet (top); an MHD jet with initial longitudinal
magnetic field configuration (middle); and an MHD jet with initial
helical magnetic field configuration (bottom), at a time $t/t_d=1.65$
($t_d=R_j/c_a \simeq 38$ years, where $c_a $ is the ambient sound
speed). The initial conditions are: $\eta=n_j/n_a=3$, $n_a=200$ cm$^{-3}$,
average ambient Mach number $M_a= v_j/c_a =24$, $v_j \simeq 398$ km
s$^{-1}$, $q_{bs}\simeq 8$ and $q_{js} \simeq 0.3$. The initial $\beta
= 8\pi p/B^2$ for the MHD cases is $\beta \simeq 1$.  The gray scale
(from minimum to maximum) is given by:  black, grey, and white.
The maximum density reached by the shell at the head
of the jets is $n_{sh}/n_a \simeq 210$ (top), $n_{sh}/n_a \simeq 153$
(middle), and $n_{sh}/n_a \simeq  160$ (bottom) (from Cerqueira, de
Gouveia Dal Pino \& Herant 1997).} 
\end{figure}

\section{Numerical Approach} 

The several models above solve the ideal-MHD conservation equations
employing different numerical techniques (the 2-D calculations have used
grid- based codes,  while the 3-D models an SPH code; see the references
above for details).  Most, simply assume a fully ionized (Hydrogen) gas
with an ideal equation of state, and a radiative cooling rate  (due to
collisional de-excitation and recombination behind the shocks) given by
the coronal cooling function tabulated for the interstellar  gas (e.g.,
Dalgarno \& McCray 72).  O'Sullivan \& Ray (2000), in particular, have
also included in their model the cooling function for the H$_2$ molecule.

Given the present uncertainties related to the real orientation and
strength of the magnetic field in protostellar jets,  three different
initial magnetic field configurations have been adopted in most
of the models with the $\beta$ parameter typically varying between
$\beta \simeq 0.1  - \infty$.  One of the adopted configurations was an
initially  constant longitudinal magnetic field parallel to the jet axis
permeating both, the jet and the ambient medium, $\vec{B} =(B_o,0,0)$
(e.g., Cerqueira et al. 1997).  Another adopted geometry was a force-free
helical magnetic field which also extends to the ambient medium (see,
e.g., Fig. 1 of Cerqueira \& de Gouveia Dal Pino 1999).  In this case, the
maximum strength of the magnetic field in the system corresponds to the
magnitude of the longitudinal component of the field at the jet axis. The
third geometry adopted was a purely toroidal magnetic field permeating
the jet only (see, e.g., Fig. 1 of Stone \& Hardee 2000).  The field
in this case is zero on the jet axis, achieves a maximum strength  at
a radial position in the jet interior (e.g., $R_m \simeq 0.9 R_j$), and
returns to zero at the jet surface.  This field corresponds to a uniform
current density inside the jet and a return current at the surface of the
jet. In the first two magnetic field configurations, the jet is assumed
to have an initially constant gas pressure ($p_j$). In the toroidal
configuration, in order to ensure an initial magnetostatic equilibrium,
the jet gas pressure has a radial profile with a maximum at the jet axis
(Stone \& Hardee 2000).

\section{Numerical Results for Steady-State Jets}

Let us first briefly review the results for steady jets, i.e., jets
which are injected with a constant velocity into the ambient medium.

Figure 1 compares 3-D simulations of two MHD jets with initial $\beta
\simeq 1$ (the middle jet has an initial constant longitudinal magnetic
field, and the bottom jet has an initial helical magnetic field)
with a pure hydrodynamical jet (top panel).  At the head, a double
shock structure develops with a forward bow shock that accelerates
the ambient medium and a reverse jet shock that decelerates the jet
material coming upstream.  As found in previous numerical HD work,
the cooling of the shocked gas at the head  results in most of the
shocked gas collecting in a dense, cold shell that eventually fragments
due to a combination of non-uniform cooling and Rayleigh-Taylor  (R-T)
instabilities (see the HD jet on the top panel).  While in the  MHD jet
with initial longitudinal field the fragmentation is still retained,
in the jet with initial helical field, the amplification of the toroidal
component of the magnetic field reduces the shock compressibility and the
growth of the R-T instability thus inhibiting the  shell-fragmentation.
This result could be an indication that the latter configuration should
dominate near the jet head as the observed jets have a clumpy structure
there (Cerqueira et al. 1997).  (See, however, how this result can be
modified in pulsed jets.)

For any geometry, the presence of magnetic fields always tend to improve
jet collimation in comparison to purely hydrodynamic calculations, due
to the amplification and reorientation of the magnetic field components
in the shocks, particularly for a helical and a toroidal geometry (e.g.,
Cerqueira et al. 1997; Cerqueira \& de Gouveia Dal Pino 1999; Frank
et al. 1998; Hardee \& Stone 2000).  Longitudinal magnetic fields of
different strengths (tested for $\beta \simeq 0.1 \, - \,10^{7}$), on
the other hand, do not strongly affect the global characteristics of  the
flow. This occur because these fields, being predominantly perpendicular
to the shock fronts, are not largely compressed in the shocks and,
therefore the hydrodynamical forces will always dominate in these
cases. However, they tend to increase order and inhibit instabilities
in the flow (Gardiner et al. 2000; O'Sullivan \& Ray 2000; Cerqueira \&
de Gouveia Dal Pino 1999).

For typical HH jet parameters, the multidimensional simulations also
indicate the development of MHD Kelvin-Helmholtz $pinch$ instabilities
along the flow (see Fig. 1), but these pinches are, in general, so
weak that it is unlikely that they could play an important  role in the
formation of the bright emission knots (e.g., Cerqueira \& de Gouveia Dal
Pino 1999; Stone \& Hardee 2000). Pure hydromagnetic pinches have been
also detected in simulations involving purely toroidal fields. These
can be particularly strong in the presence of intense fields, for example,
 close to
the jet source (e.g., Frank et al. 1998).

\begin{figure*}
  \includegraphics[width=\textwidth]{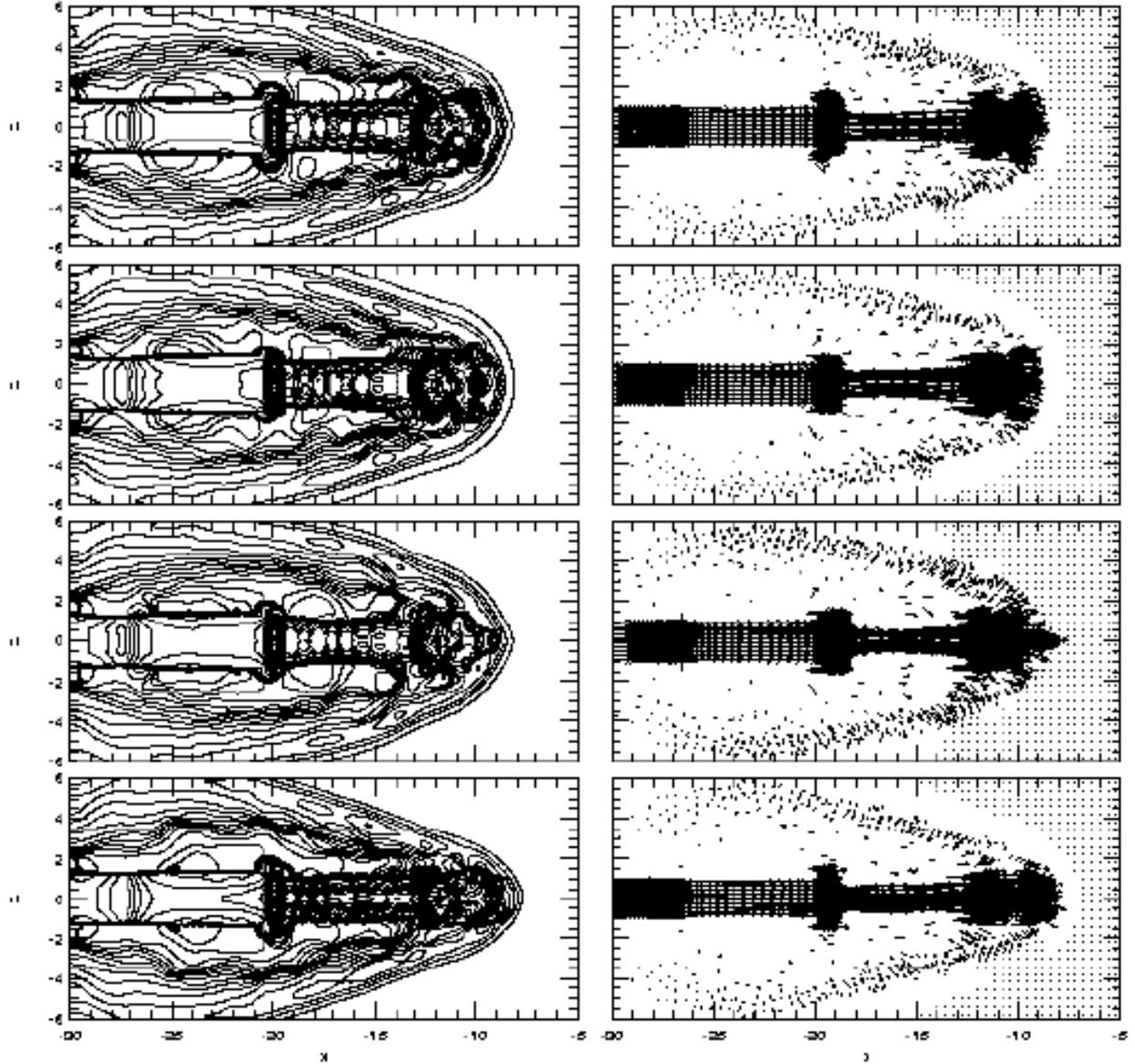}
  \caption{ Midplane density contours (left) and distribution of
velocity vectors (right) for (from top to bottom): a purely hydrodynamical
pulsed jet; an MHD jet with initial longitudinal magnetic field;
an MHD jet with initial helical magnetic field; and an MHD jet with
initial toroidal magnetic field, at a time $t/t_d=1.75$. The initial
conditions are: average ambient Mach number $M_a= v_j/c_a \simeq 15$,
$v_0 =v_j[1+A{\rm sin}(2\pi t / P)]$, with $v_j \simeq 250$ km s$^{-1}$,
$A=0.25 v_j$ and $P=0.54 t_d$, $\eta= n_j/n_a =5$, $\beta =  8\pi p/B^2
= \infty$ for the HD model, and maximum $\beta \simeq 1$ for the MHD
models. The maximum density in each model (from top to bottom) is: 119,
115, 124 and 175 (1 c.u. $= 200$ cm$^{-3}$) (from Cerqueira \& de Gouveia Dal
Pino 2001a) .}
\end{figure*}

\section{Numerical Results for Pulsed Jets}

The bright emission knots immersed in the HH jets are one of their most
prominent features. They frequently exhibit a bow shock morphology and
high spatial velocity (e.g., Bally \& Reiputh 2001, these proceedings)
which indicate that they are shocks that arise from the steepening of
velocity (and/or density) fluctuations in the underlying, supersonic
outflow.  Strong support for this conjecture has been given by theoretical
studies which have confirmed that traveling shocks created in this way
reproduce the essential properties of the observed knots (e.g., Raga et
al. 1990; Kofman \& Raga 1992; Stone \& Norman 1993; de Gouveia Dal Pino
\& Benz 1994; V\"olker et al. 1999; Raga \& Cant\'o 1998; de Gouveia
Dal Pino 2001).

To simulate a pulsing jet, it is usually applied a sinusoidal velocity
variation with time at the inlet: $v_o(t)= v_j [1 + A\cdot{\rm sen}({{2
\pi}\over P} \cdot t)]$, where $v_j$ is the mean jet speed, $A$ the
amplitude of the velocity oscillation, and $P$ the oscillation period.

\begin{figure}
  \includegraphics[width=\columnwidth]{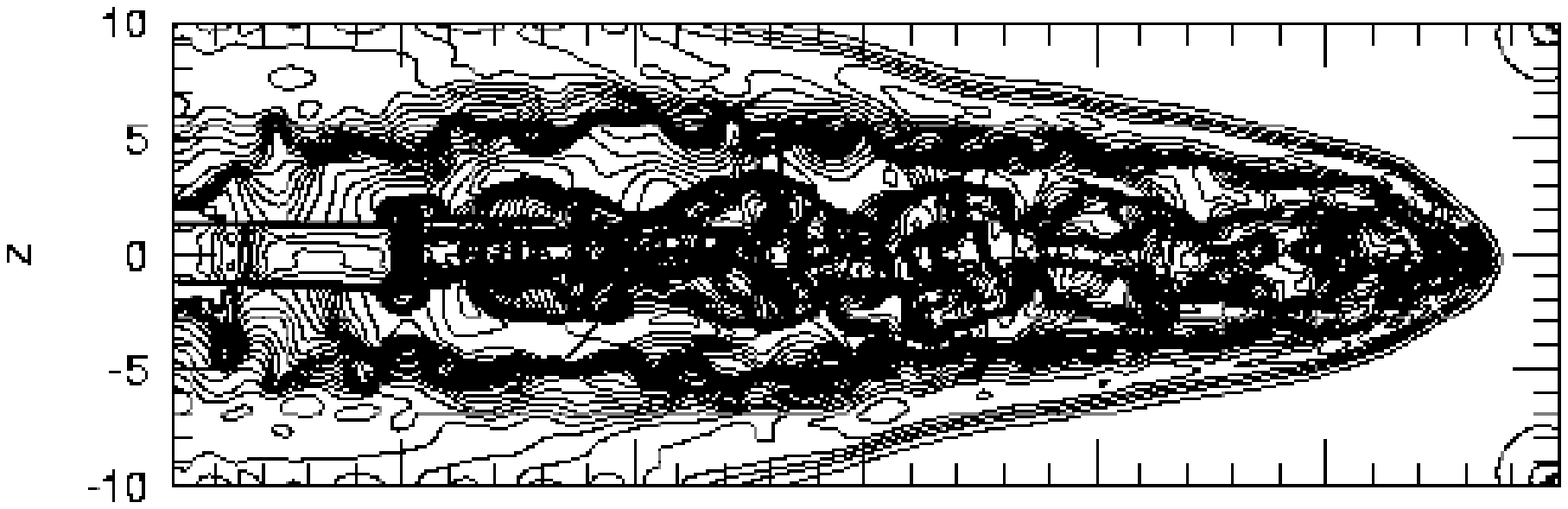}
  \caption{ Midplane density contours  for the MHD jet of Fig. 2 with initial 
helical magnetic field at a time $t/t_d=5$ (from Cerqueira \& de Gouveia Dal 
Pino 2001b).}
\end{figure}

2-D axisymmetric calculations of pulsed jets show that the presence
of toroidal magnetic fields may cause dramatic effects on pulsed jets
depending on the initial conditions. For example, Stone \& Hardee (2000)
have found that the radial hoop stresses due to the toroidal field confine
the shocked jet material in the internal pulses resulting in higher
densities in the pulses which are strongly peaked towards the jet axis in
comparison to purely hydrodynamic calculations.  O'Sullivan \& Ray (2000),
on the other hand, carrying out 2-D calculations of jets with initially
lower densities (corresponding to a density ratio between the jet and
the ambient medium $\eta$ = 1), and which are highly overpressurized
with respect to the ambient medium, find the development of more complex
cocoons around the beams and the formation of crossing shocks which help
to  refocus the beam  and the internal pulses. In particular, they find
that the  hoop stresses associated with toroidal fields can cause a high
degree of H ionization and H$_2$  dissociation, and the $disruption$
of the internal knots, even for $\beta \simeq $1.  Also, in these 2-D
models, $nose$ $cones$ are found to develop at the head of magnetized
jets with toroidal components. These features are extended narrow plugs
that form as most of the  shocked cooled  gas between the jet shock and
the bow shock at the head is prevented from escaping sideways into the
cocoon (and confined towards the jet axis) by the toroidal field.

\begin{figure}
\vskip -8.0truecm
  \includegraphics[width=\columnwidth]{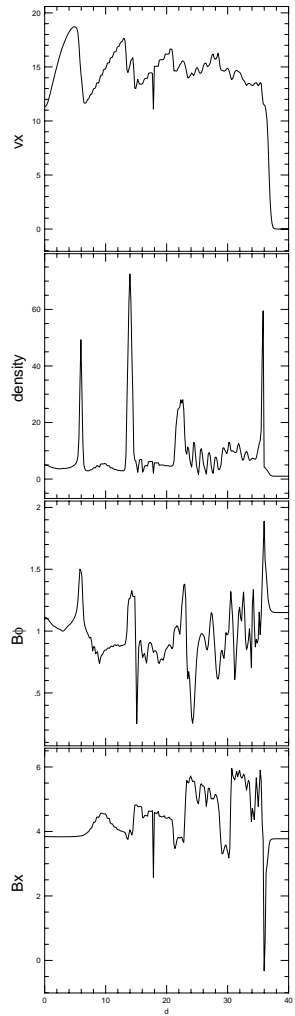}
\vskip 5.65truecm
  \caption{ From top to bottom: velocity, density, toroidal
magnetic field component, and longitudinal magnetic field profiles along
the beam axis for the jet of Fig. 2 with initial helical magnetic field.
Time displayed $t/t_d \simeq 3 $.  (from Cerqueira \& de Gouveia Dal Pino
2001a). }
  \end{figure}

Now, 3-D  MHD calculations of pulsed jets reveal significant changes
relative to the 2-D models. The morphological features tend to be
generally smoothed out in 3-D, and the differences that arise in
2-D calculations with distinct field geometries seem to diminish in
the 3-D models (Cerqueira \& de Gouveia Dal Pino 2001a, b).  As an
example, Figure 2 displays the midplane density contours (left) and
the velocity field distribution (right), for four supermagnetosonic,
radiatively cooling, $pulsed$  jets in their early evolution, after
they have propagated over a distance $\approx 22 R_j$ (where $R_j$ is
the jet radius).  The top jet is purely hydrodynamical (HD), the second
jet has an initial constant longitudinal magnetic field configuration;
the third an initial helical magnetic field, and the bottom an initial
toroidal field.  At the time depicted, the leading $working$ $surface$
at the jet head is followed by 2 other features and a third pulse is just
entering into the system. Like the leading working surface, each internal
feature consists of a double-shock structure, an upstream reverse shock
that decelerates the high velocity material entering the pulse, and a
downstream forward shock  sweeping up the low velocity material ahead of
the  pulse.  At this time, none the internal working surfaces (IWS) or
knots, has reached the jet head yet. Later on, however, one by one of them
reach the terminal working surface and is disrupted by the impact. Their
debris are partially deposited into the cocoon thus drastically changing
the head morphology and providing a complex fragmented structure, as
in Figure 3, where the jet of Figure 2 with initial helical field is
depicted at a later time of its evolution. No nose cones have developed,
nor even in the jet with initial toroidal field. We speculate that their
absence in the 3-D calculations could be, in part, explained by the
fact that, as the magnetic forces are intrinsically three-dimensional,
they cause part of  the material to be deflected in a third direction,
therefore, smoothing out the strong focusing of the shocked material that
is otherwise detected in the 2-D calculations (Cerqueira \& de Gouveia
Dal Pino 2001a; Stone \& Hardee 2000 have also argued that nose cones
should be unstable in 3-dimensions).

Figure 2 also indicates that the overall morphology of the 3-D pulsed jet
is not very much affected by the presence of the different magnetic field
configurations when compared to the purely HD calculation. Instead, the
distinct $\bf{B}$-profiles  tend to alter only the  detailed structure
behind the shocks at the head and internal knots, particularly in the
helical and purely toroidal cases.  Like in steady calculations, the pure
HD jet exhibits a cold dense shell at the head, formed by the cooling
of the shock-heated material, that has  become also R-T unstable and
separated into clumps. Again, these can also be identified in the shell
of the MHD jet with initial longitudinal field, but not in the helical or
the toroidal cases.  The 3-D simulations also evidence, as in previous
work (e.g., de Gouveia Dal Pino \& Benz 1993), that the density of the
shell (and IWSs) undergoes oscillations with time (with a period of the
order of the cooling time behind the shocks) which are caused by global
thermal instabilities of the radiative shocks.  The profile and maximum
amplitude of these oscillations are particularly affected by the presence
of toroidal field components (Cerqueira \& de Gouveia Dal Pino 2001b).

\begin{figure}
\vskip -2truecm
  \includegraphics[width=\columnwidth]{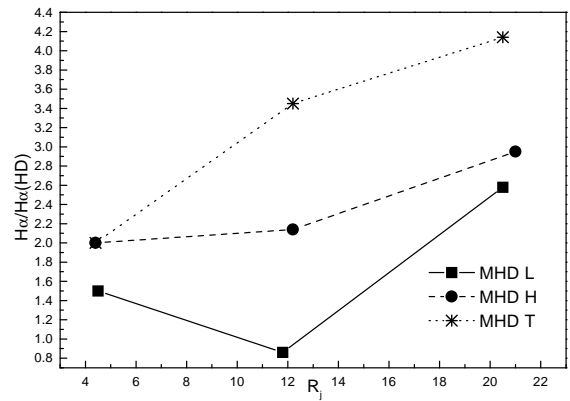}
\vskip -4.0truecm
  \caption{ The ratio between the H$_{\alpha}$ intensity along
the jet axis within the IWSs for the different magnetized jets of Fig. 2
and the the H$_{\alpha}$ intensity of the purely hydrodynamical jet.  (from 
Cerqueira \& de Gouveia Dal Pino 2001a). }
  \end{figure}

Some wiggling can be detected in the jet axis in the 3-D MHD simulation
with initial helical field of Figure 3. This is caused by the $kink$ mode
of the K-H instability and was also detected (but in smaller intense)
in the pure hydrodynamical counterpart. This could provide a potential
explanation for the wiggling features often detected in HH jets.

Several physical quantities along the beam axis of the jet of Figure 3
with initial helical magnetic field are depicted in Figure 4, at a time
$t/t_d \simeq 3$, when then two more pulses have entered the system and
developed new IWSs, and the first IWS has already merged  with the jet
head. The initial sinusoidal velocity profile applied to the flow at
the inlet steepens  into the sawtooth pattern (top panel) familiar from
earlier studies of oscillating jets (e.g., Raga \& Kofman 1992) as the
faster material catches up with the slower, upstream gas in each pulse.
The sharp density spikes within each IWS (second panel) are correlated
with the toroidal component of the magnetic field. The third panel  shows
that this component $B_{\phi}$ (third panel) sharpens within the knots
and rarefies between them, while the longitudinal component, $B_{//}$
(forth panel) is stronger between the knots (Cerqueira \& de Gouveia
Dal Pino 2001a).  This result is in agreement with 2-D calculations
(Gardiner \& Frank 2000; O'Sullivan \& Ray 2000) and could, in principle,
be checked observationally as a diagnostic of helical geometries. Helical
fields are specially attractive for protostellar jets as they are the
predicted geometry from magneto-centrifugally launched models (Gardiner \&
Frank 2000; Lerry \& Frank 2000).

Although the distinct $\bf{B}$-geometries do not seem to significantly
affect the global characteristics of the 3-D flows, they can modify
the details of the emission structure, particularly for the helical and
toroidal cases. In particular, Figure 5 shows the H$_{\alpha}$ intensity
along the jet axis evaluated within the IWSs for the different magnetized
jets of Figure 2 which are compared to the H$_{\alpha}$ intensity of
the baseline hydrodynamical jet.  The H$_{\alpha}$ intensity is observed
to be strong behind the knots of the protostellar jets and is predicted
to have a strong dependence with the shock speed $v_s$, $ I_{{\rm H}{\alpha}}
\propto \rho_{d} v_s^{3.8} $ where $\rho_{d}$ is the downstream pre-shock
density (e.g., Raga \& Cant\'o 1998). We have used this relation and the
results of the simulated jets of Figure 2 to evaluate the intensity ratios
of Figure 4.  We note, as expected, that the intensity ratio does not
differ much from unity in the case of the magnetized jet with initial
longitudinal field, while for the helical and toroidal field cases,
the H$_{\alpha}$ intensity can be about four times larger (for $\beta
\simeq 1$ jets) than that of the HD case.

\section{Conclusions}

The effects of magnetic fields are dependent on both the field-geometry
and intensity (which, unfortunately, are still poorly determined from
observations). The  presence of a helical or a toroidal field tends to
affect more the characteristics of the fluid, compared to the purely HD
calculation, than a longitudinal field.  However, the relative differences
which are detected in 2-D simulations  involving distinct magnetic field
geometries (e.g., Stone \& Hardee 2000; O'Sullivan \& Ray 2000), seem to
decrease in the 3-D calculations (Cerqueira \& de Gouveia Dal Pino 2001a,
b).  In particular, we have found that features, like the  nose cones,
that often develop at the jet head in 2-D calculations involving toroidal
magnetic fields, are absent in the 3-D models (Cerqueira \& de Gouveia
Dal Pino 1999, 2001a, b), a result which is consistent with observations which
show no direct evidence for nose cones at the head of protostellar jets.

In 3-D calculations, magnetic fields which are initially nearly  in
equipartition with the gas tend to affect essentially  the detailed
structure behind the shocks at the head and internal knots, mainly
for the helical and toroidal topologies. In such cases, the H$_\alpha$
emissivity behind the internal knots can increase by a factor up to four
relative to that in  the purely hydrodynamical jet
(Cerqueira \& de Gouveia Dal Pino 2001a).

Further 3-D MHD studies are still required as the detailed structure and
emission properties of the jets seem to be sensitive to multidimensional
effects when magnetic forces are present. Also obvious is the need for
further observations and polarization mapping of star formation regions,
for a real comprehension of the magnetic field structure of protostellar
jets.

\acknowledgements 
Financial support for this research has been provided by the Brazilian
agencies FAPESP and CNPq.


\end{document}